\makeatletter
\@ifundefined{@parse@version@dash}{%
\def\@parse@version#1{\@parse@version@0#1}
\def\@parse@version@#1/#2/#3#4#5\@nil{%
\@parse@version@dash#1-#2-#3#4\@nil}
\def\@parse@version@dash#1-#2-#3#4#5\@nil{%
  \if\relax#2\relax\else#1\fi#2#3#4 }
}{}
\makeatother

\documentclass[%
 reprint,
 amsmath,amssymb,
 aps,
 prl,
]{revtex4-1}

\usepackage{graphicx}
\usepackage{dcolumn}
\usepackage{bm}
\usepackage{lineno}
\usepackage{soul}
\usepackage[normalem]{ulem}

\usepackage{color}
\usepackage{xcolor}

\newcommand {\rootsNN}  {\ensuremath{\sqrt{s_{_{\rm{NN}}}}}}

\usepackage{color}

\usepackage{multirow}

\begin{document}

\title{Elliptic Flow of Multi-Strange Hadrons in Au+Au Collisions at $\sqrt{s_{NN}}$ = 7.7–19.6 GeV}
\author{The STAR Collaboration}
\date{\today}
\begin{abstract}
Multi-strange hadrons ($\phi$, $\Xi$, and $\Omega$), owing to their small hadronic interaction cross sections, provide a sensitive probe of partonic collectivity with minimal distortion from late-stage hadronic rescattering. Using high-statistics data from the second phase of the STAR Beam Energy Scan program, we present precision measurements of elliptic flow ($v_2$) for multi-strange hadrons, together with other identified hadrons ($\pi^{\pm}, K^{\pm}$, $K_{S}^{0}$, $p$, and $\Lambda$) for comparison, in Au+Au collisions at $\sqrt{s_{NN}}$ = 7.7--19.6 GeV. Number-of-constituent-quark (NCQ) scaling is examined separately for particles and antiparticles. The NCQ-scaled $v_2$ of multi-strange hadrons follows the scaling observed for other identified species, with ratios relative to $K_{S}^{0}$ close to unity and exhibiting weak beam-energy dependence. The persistence of NCQ scaling for these weakly interacting probes provides strong evidence that collective flow is predominantly developed in the partonic phase, persisting down to $\sqrt{s_{NN}}$ = 7.7 GeV. The similar particle-antiparticle $v_2$ splittings observed for $p$, $\Lambda$, and $\Xi$, despite their markedly different hadronic cross sections, indicate that the splitting depends weakly on hadronic interaction strength. This suggests that late-stage hadronic interactions are subdominant and that the observed splitting primarily reflects dynamics established in the early, partonic stage. 

\end{abstract}

\maketitle
{\bf Introduction.} 
Understanding the relevant degrees of freedom in strongly interacting matter at finite baryon density is a central objective of the Beam Energy Scan (BES) program at the Relativistic Heavy Ion Collider (RHIC). By varying the collision energy, the BES program explores a broad region of the quantum chromodynamics (QCD) phase diagram in terms of the temperature and baryon chemical potential, where the dominant degrees of freedom are expected to vary between partonic and hadronic~\cite{STAR:2010vob}. A key question in this program is the beam energy dependence of collective flow and whether it primarily reflects early partonic dynamics or late-stage hadronic interactions. 

The azimuthal anisotropy of particle emission can be characterized by the Fourier coefficients $v_n$ of the final-state hadron distribution~\cite{Poskanzer:1998yz,Voloshin:2008dg},
\begin{equation}
\frac{dN}{d\phi} \propto 1 + 2\sum_{n=1}^{\infty} v_n \cos[n(\phi - \Psi_{\rm n})],
\end{equation}
where $\phi$ denotes the azimuthal angle of the particle, and $\Psi_n$ is the event-plane angle associated with the $n^{\mathrm{th}}$-order harmonic flow.
The elliptic flow coefficient $v_2$ is generated predominantly during the early stages of the collision, although subsequent hadronic rescattering can modify its magnitude. 
It therefore provides sensitivity to both the early partonic phase and the later hadronic evolution.

At top RHIC and Large Hadron Collider (LHC) energies, the approximate number of constituent quarks (NCQ) scaling of $v_2$ has been widely interpreted as evidence that collective motion is established at the partonic level, followed by hadronization via quark coalescence~\cite{Adams:2005dq,PHENIX:2006dpn,Abelev:2007rw,Abelev:2007qg,STAR:2015gge,Adamczyk:2017xur,ALICE:2014wao,CMS:2018loe,Chen:2026gka}.
In the fixed-target program of the STAR experiment ($\sqrt{s_{NN}} = 3$--4.5 GeV), which probes matter at low temperature and high baryon density, NCQ scaling is observed to break down at $\sqrt{s_{NN}} \leq 3.2$ GeV and gradually re-emerges toward $\sqrt{s_{NN}} \geq 4.5$ GeV for light-flavor hadrons, such as $\pi^{\pm}$, $K^{\pm}$, $K_S^{0}$, $p$, and $\Lambda$~\cite{STAR:2025owm}. Between the top RHIC/LHC energies and the fixed-target regime lies an intermediate beam-energy range in which high-precision measurements of multi-strange hadron elliptic flow, crucial for probing partonic collectivity, are still lacking. Consequently, the evolution of NCQ scaling, including the energy threshold for the onset of dominant partonic collectivity in this region, remains an open question. This energy scale is therefore essential for mapping the QCD phase diagram.
Multi-strange hadrons such as $\phi$, $\Xi$, and $\Omega$, provide uniquely clean probes of this evolution. 

Furthermore, the typical hierarchy of hadronic interaction cross sections in the low-energy regime relevant for hadronic rescattering~\cite{Smith:1997xu,Chung:1997mp,Ferrari:1959,Gal:2016boi,Crede:2024hur,ParticleDataGroup:2024cfk},
$\sigma_{\phi\text{-hadron}} \lesssim 2~\mathrm{mb}$,
$\sigma_{\Omega\text{-N}} \lesssim 5~\mathrm{mb}$,
$\sigma_{\Xi\text{-N}} \sim 10$--$15~\mathrm{mb}$,
$\sigma_{\Lambda\text{-N}} \sim 20$--$30~\mathrm{mb}$, and
$\sigma_{\text{N-N}} \sim 32~\mathrm{mb}$,
implies a systematic reduction of hadronic rescattering from nucleons to $\Lambda$, $\Xi$, $\Omega$ and $\phi$. Among these, the $\phi$, $\Xi$, and $\Omega$ are particularly weakly coupled to the hadronic medium and thus serve as sensitive messengers of partonic flow. A common NCQ scaling behavior shared by multi-strange and light-flavor hadrons would therefore signal that collective motion is established prior to hadronic freeze-out.

At $\sqrt{s_{NN}} < 39$ GeV, stronger baryon stopping produces a baryon-rich medium with substantial net-baryon density, in which sizable particle-antiparticle $v_2$ splittings have been attributed to baryon transport and mean-field effects, with possible contributions from hadronic interactions~\cite{Steinheimer:2012bn,Xu:2012gf,Xu:2013sta,Liu:2019ags,Dunlop:2011cf}. Because multi-strange hadrons couple only weakly to the hadronic phase, the particle–antiparticle splittings of $p$, $\Lambda$, and $\Xi$ provide a differential test of the underlying dynamics. If similar splittings persist despite large differences in hadronic cross sections, this would favor an origin in early-stage baryon-related dynamics, presumably including baryon transport and possible mean-field effects in the finite-density medium, rather than late-stage hadronic rescattering.

\begin{figure*}[htbp]
  \centering
  \includegraphics[width=1.0\textwidth]{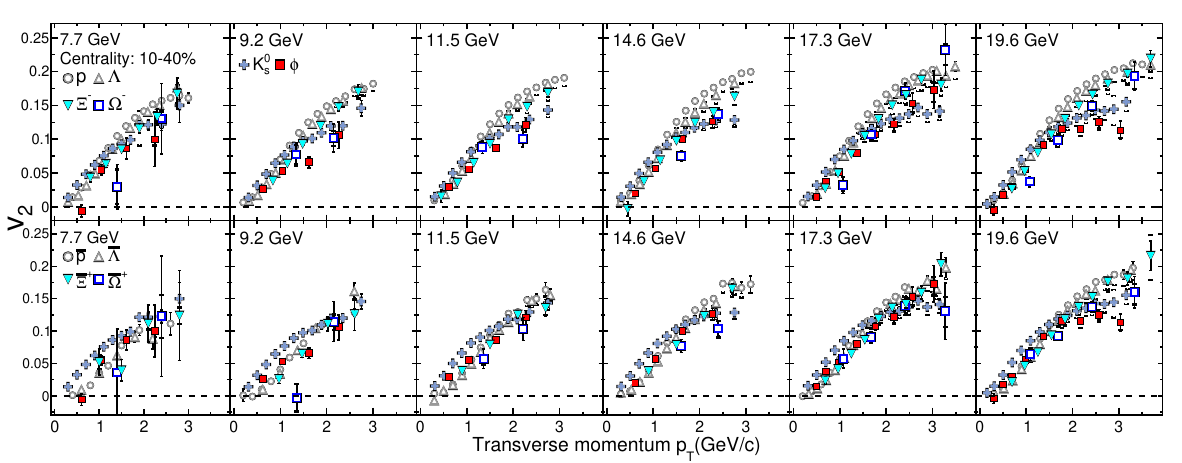}
  \caption{Elliptic flow $v_2$ as a function of $p_T$ in 10--40\% centrality Au+Au collisions at $\rootsNN = 7.7$--19.6~GeV is shown for multi-strange hadrons ($\phi$, $\Xi$, and $\Omega$), protons, $\Lambda$, and $K_S^0$. 
The top panels show $v_2(p_T)$ for particles, while the bottom panels show that for antiparticles. 
Error bars and brackets represent statistical and systematic uncertainties, respectively.}     
  \label{Fig:fig1}
\end{figure*}

\begin{figure*}[htbp!]
    \centering
    \includegraphics[width=1.0\textwidth]{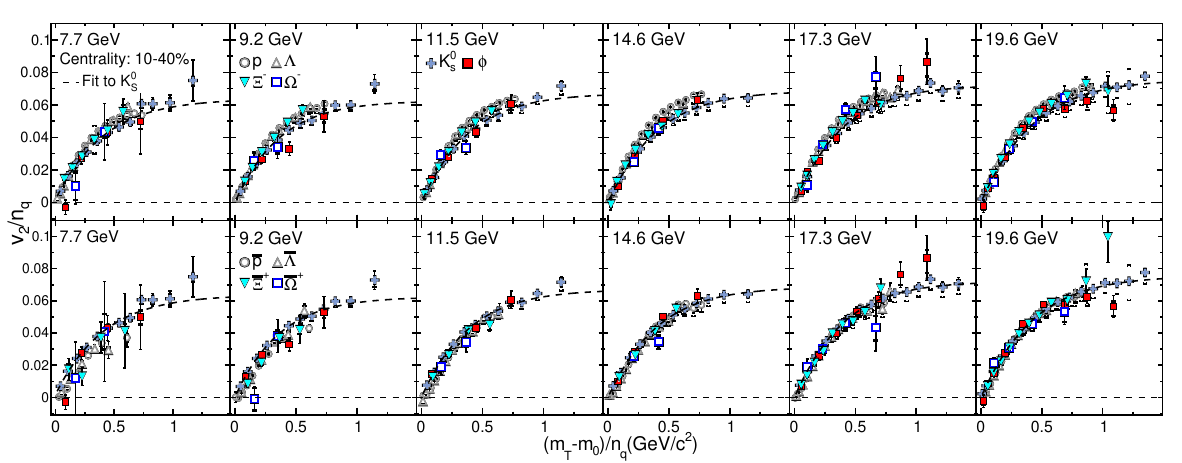}
    \caption{The NCQ-scaled elliptic flow, $v_2/n_q$ versus $(m_T - m_0)/n_q$, in 10–40\% centrality Au+Au collisions at $\rootsNN$ = 7.7-19.6 GeV is shown for multi-strange hadrons ($\phi$, $\Xi$, and $\Omega$), protons, $\Lambda$, and $K_S^0$. The dashed lines represent fits to the $K_S^0$ data ~\cite{Dong_2004}. The top panels present the NCQ scaling behavior of particles, while the bottom panels present that of antiparticles. Error bars and brackets represent statistical and systematic uncertainties, respectively.}
    \label{Fig:fig2}
\end{figure*}

\begin{figure*}[htbp!]
    \centering
    \includegraphics[width=0.76\textwidth]{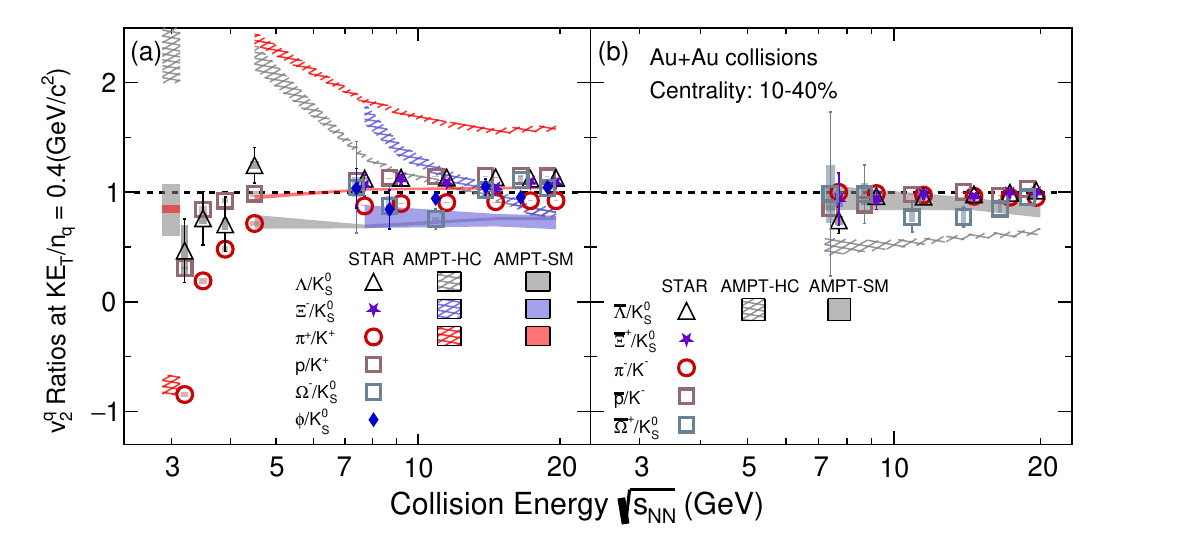}
    \caption{Energy dependence of NCQ-scaled $v_2/n_q$ ratios at $(m_T - m_0)/n_q = 0.4$~GeV/$c^2$ in 10--40\% centrality Au+Au collisions. 
(a) Ratios for particles: $\Lambda/K_S^0$, $\Xi^-/K_S^0$, $\pi^+/K^+$, $p/K^+$, $\Omega^-/K_S^0$, and $\phi/K_S^0$. The results for 3.2-4.5 GeV are taken from Ref.~\cite{STAR:2025owm}.
(b) Ratios for antiparticles: $\bar{\Lambda}/K_S^0$, $\bar{\Xi}^+/K_S^0$, $\pi^-/K^-$, $\bar{p}/K^-$, and $\bar{\Omega}^+/K_S^0$. 
Colored bands show results from AMPT-SM~\cite{He:2017tla} and AMPT-HC~\cite{YONG2021136521} model calculations. For clarity, the AMPT-HC result of $\Lambda/K_S^0$ at 3 GeV are scaled down by a factor of 3. Error bars and bands represent statistical and systematic uncertainties.}

    \label{Fig:fig3}
\end{figure*}

\begin{figure}[htbp!]
    \includegraphics[width=1.0\columnwidth]
    {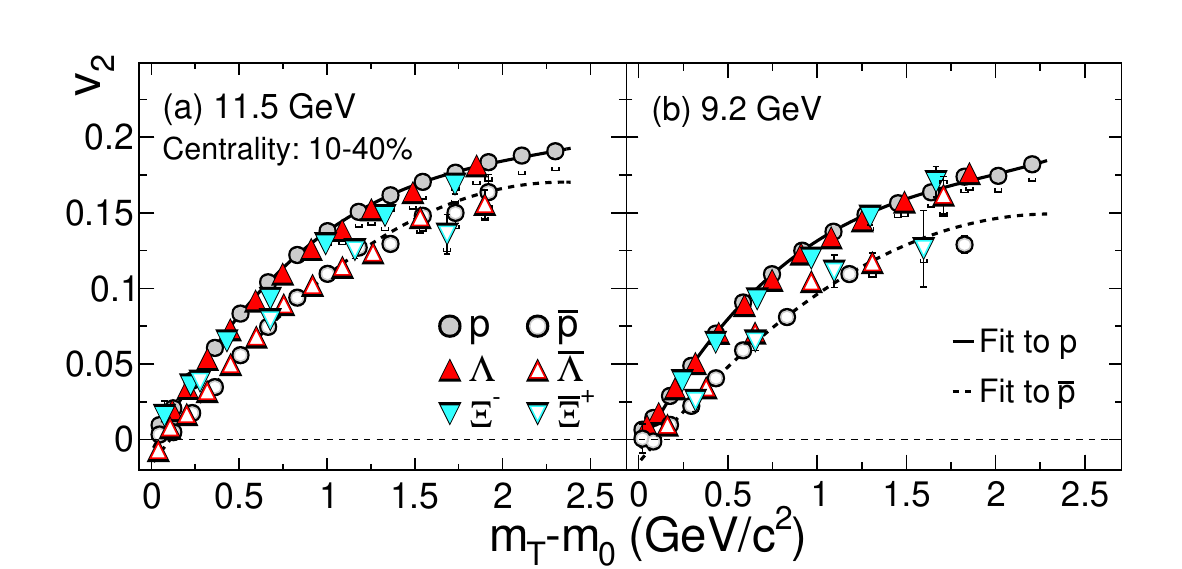}
    \caption{Elliptic flow $v_2$ as a function of $(m_T - m_0)$ for selected particles and antiparticles in 10--40\% centrality Au+Au collisions at $\rootsNN =$ 11.5 and 9.2~GeV. 
Panels (a) and (b) show the results at 11.5 and 9.2 GeV, respectively. Particles and antiparticles are represented by filled and open markers, respectively. Error bars indicate statistical uncertainties, and brackets represent systematic uncertainties. The solid and dashed lines represent fits to the proton and antiproton data with third-order polynomial functions.}

    \label{Fig:fig4}
\end{figure}

{\bf Analysis.}
The second phase of the STAR Beam Energy Scan program (BES-II), with substantially increased statistics and upgraded detectors, enables an unprecedented high-precision study of multi-strange hadron flow in Au+Au collisions at $\sqrt{s_{NN}} = 7.7$--19.6 GeV~\cite{Yang:2017llt,Wang:2023vqi,Adams:2019fpo, Chen:2024BESreview}. In this letter, we report measurements of the elliptic flow for multi-strange hadrons ($\phi$, $\Xi$, and $\Omega$) with light-flavor hadrons ($\pi^{\pm}$, $K^{\pm}$, $K_S^{0}$, $p$, and $\Lambda$) included for comparison, in Au+Au collisions at $\rootsNN = 7.7$, 9.2, 11.5, 14.6, 17.3, and 19.6 GeV, using 96, 246, 334, 375, 348, and 716 million minimum-bias events after good event selection, respectively. These data samples are typically about 20–30 times larger than those in the first phase of STAR’s Beam Energy Scan program (BES-I).
For each event, the primary vertex along the beam axis is required to lie within ±145 cm of the Time Projection Chamber (TPC) center, and the radial vertex position must be smaller than 2 cm. Collision centrality is determined from charged-particle multiplicity in pseudorapidities $|\eta| < 0.5$ via a Monte Carlo Glauber simulation~\cite{Miller:2007ri}. 
Tracks are required to have at least 15 TPC fit points ($\rm nHitsFit$), and primary particles are required to have a distance of closest approach (DCA) to the primary vertex of smaller than 3 cm.

Long-lived charged hadrons ($\pi^{\pm}$, $K^{\pm}$, $p$, and $\bar{p}$) are identified using the mean specific energy loss $\langle dE/dx \rangle$ measured in the TPC and the mass squared ($m^{2}$) information provided by the Time-of-Flight (TOF) detector~\cite{STAR:2013ayu}. 
Short-lived particles ($\phi$, $K^0_S$, $\Lambda$, $\Xi$, and $\Omega$) are reconstructed via their decay daughters: $K^0_S$, $\Lambda$, $\Xi$, $\Omega$ with the Kalman Filter Particle method~\cite{Kisel:2018nvd,Banerjee:2020iab}, and $\phi$ mesons via invariant mass analysis of $K^+K^-$ pairs~\cite{STAR:2008bgi}. The invariant mass distributions of $\phi$, $K^0_S$, $\Lambda$, $\Xi$, and $\Omega$ are provided in the Supplemental Material~\cite{SupplementalMaterial}.

The event plane is reconstructed from TPC tracks within $|\eta|<1.5$ using the $\eta$-subevent method, where the full event is divided into two symmetric subevents separated by a 0.1 $\eta$ gap. 
Event-by-event recentering and shifting calibrations are applied to suppress detector acceptance effects~\cite{Voloshin:2008dg}, and particles are correlated with the opposite $\eta$-subevent plane to suppress non-flow effects, with positive and negative $\eta$ treated in a cross manner; the resulting $v_2$ is denoted as $v_2\{\eta_{\rm sub}\}$. 
The event-plane resolution is evaluated in each centrality bin from the correlation between the event-plane angles of the two $\eta$ subevents and is used to correct the measured $v_{2}$ in wide centrality bins~\cite{Masui:2012zh}. 
The inner Time Projection Chamber~\cite{SHEN201890} upgrade results in an approximately 10\% improvement in the event-plane resolution relative to that from BES-I.

Systematic uncertainties are evaluated by varying track quality and particle identification selections. 
For primary hadrons, DCA, $\rm nHitsFit$, and the number of standard deviations between the measured and expected energy loss ($|n\sigma|$) are varied. For $\phi$ mesons and weakly decaying particles, the daughter-track selection criteria are varied, with DCA used for $\phi$ reconstruction, topological cuts for weakly decaying particles, and $\rm nHitsFit$ and $|n\sigma|$ applied to both.
Two independent variations per source are applied, and the largest accepted deviation is taken as the systematic uncertainty using a Barlow check~\cite{Barlow:2002yb}. 
Residual non-flow effects are estimated from the difference between $v_2\{\eta_{\rm sub}\}$ ($\eta_{\rm gap}\ge0.1$) and $v_2\{4\}$~\cite{Bilandzic:2010jr}, serving as a conservative one-sided lower uncertainty, with values in the range of 4-9\%.
All contributions are combined in quadrature. Typical uncertainties are less than 10\% for most hadrons, and larger at low $p_T$ for antiprotons and multi-strange baryons (up to 10–30\%).

{\bf Results.}
Figure~\ref{Fig:fig1} presents $v_2$ as a function of $p_T$ for ${K}^{0}_{S}$, $p$, $\bar{p}$, ${\phi}$, ${\Lambda}$, $\bar{\Lambda}$, ${\Xi}^{-}$, $\bar{\Xi}^{+}$, ${\Omega}^{-}$, and $\bar{\Omega}^{+}$ in 10–40\% central Au+Au collisions at $\sqrt{s_{NN}} = 7.7$, 9.2, 11.5, 14.6, 17.3, and 19.6 GeV. Additional results for $\pi^{\pm}$ and $K^{\pm}$ are provided in the Supplemental Material~\cite{SupplementalMaterial} for comparison. The $v_2(p_T)$ results for particles and antiparticles are shown in the top and bottom panels, respectively. These new results are consistent with the previous BES-I results, but with much smaller uncertainties~\cite{STAR:2013cow}.
At all beam energies, a clear mass ordering is observed at $p_T < 1.5~\mathrm{GeV}/c$, where heavier hadrons exhibit smaller $v_2$ than lighter ones, consistent with hydrodynamic expectations for a collectively expanding medium~\cite{Huovinen:2001cy}. At intermediate $p_T$ ($1.5 < p_T < 3.0~\mathrm{GeV}/c$), a pronounced baryon–meson splitting emerges, including for the multi-strange hadrons: baryons ($\Lambda$, $\Xi$, $\Omega$) display systematically larger $v_2$ than mesons ($K^{0}_{S}$, $\phi$). Owing to large statistical uncertainties, this phenomenon is not evident at 7.7 GeV, particularly for antiparticles. The observation of such splitting for $\phi$, $\Xi$, and $\Omega$, which have significantly smaller hadronic interaction cross sections, suggests that late-stage hadronic rescattering alone may not be sufficient to account for this feature. Instead, it supports an interpretation in terms of quark coalescence, in which hadrons inherit the collective flow of their constituent quarks. These observations indicate the dominance of partonic degrees of freedom in Au+Au collisions down to $\sqrt{s_{NN}} = 7.7~\mathrm{GeV}$~\cite{Moln_r_2003,STAR:2003wqp,STAR:2007mum}.

Motivated by the particle–antiparticle $v_2$ differences observed in the RHIC BES-I program~\cite{STAR:2013cow}, we examine NCQ scaling separately for selected hadrons and their antiparticles. 
In Fig.~\ref{Fig:fig2}, $v_2$ scaled by the number of constituent quarks ($n_q$) is plotted as a function of the scaled transverse kinetic energy $(m_T - m_0)/n_q$ for 10–40\% centrality Au+Au collisions at $\rootsNN$ = 7.7–19.6 GeV, where $m_T = \sqrt{p^{2}_{T}+m^{2}_0}$, $m_0$ is the rest mass of the particle. A fit to the $K^{0}_{S}$ data at each energy is shown as the dashed line~\cite{Dong_2004}. The top panels show particles and the bottom panels antiparticles. 
For completeness, the NCQ scaling results based on $p_T/n_q$ are provided in the Supplemental Material~\cite{SupplementalMaterial}.
Approximate NCQ scaling is observed for both particles and antiparticles, including $\phi$ mesons and multi-strange hadrons ($\Xi$ and $\Omega$), which have relatively small hadronic cross sections and are expected to be sensitive primarily to the early stages of the collision.  

During BES-I, possible deviations from NCQ scaling in $v_2$ for the $\phi$ and $\Xi$ were reported at $\sqrt{s_{NN}} = 7.7$ GeV, but limited statistics prevented firm conclusions~\cite{STAR:2013cow,STAR:2013ayu,STAR:2015rxv,STAR:2012och}. The higher-precision BES-II results now demonstrate that NCQ scaling persists down to 7.7 GeV, indicating that partonic collectivity is established at this beam energy.

To quantify the energy dependence of NCQ scaling, Fig.~\ref{Fig:fig3} shows the ratios of $v_2/n_q$ for selected hadrons at $(m_T - m_0)/n_q = 0.4$ GeV, which corresponds to the intermediate $p_T$ region and balances statistical precision for $\phi$ and $\Xi$ hadrons. 
Figure~\ref{Fig:fig3}(a) shows $v_2/n_q$ ratios of $\Lambda$, $\Xi^-$, $\pi^+$, $p$, $\Omega^-$ and $\phi$ mesons relative to kaons, while Fig.~\ref{Fig:fig3}(b) shows the corresponding ratios for $\bar{\Lambda}$, $\bar{\Xi}^+$, $\pi^-$, $\bar{p}$ and $\bar{\Omega}^+$. The results for 3.2-4.5 GeV are from Ref.~\cite{STAR:2025owm}.
In the quark coalescence picture, two or three constituent quarks form a meson or baryon, so NCQ scaling remains an approximate law, neglecting effects from strong interactions, quark mass differences, and other correlations~\cite{Voloshin:2005up}.

Significant deviations from unity are observed only at $\sqrt{s_{NN}} = 3.2$--3.5 GeV, possibly reflecting a change in the underlying degrees of freedom near the QCD phase transition in this energy region, transported quark effects~\cite{Dunlop:2011cf}, or variations in net-baryon density~\cite{Denicol:2018wdp}.
At $\sqrt{s_{NN}} = 7.7$–19.6 GeV, all ratios—including $\phi/K^0_S$ and $\Xi/K^0_S$—remain near unity for both particles and antiparticles, with little energy dependence, providing clear evidence of partonic collectivity. Colored bands in Fig.~\ref{Fig:fig3} show calculations from the A Multi-Phase Transport model: Hadron Cascade (AMPT-HC) and String Melting (AMPT-SM) model calculations~\cite{He:2017tla, YONG2021136521}. AMPT-SM, which includes partonic interactions, approaches unity more closely and shows weaker beam-energy dependence than AMPT-HC, highlighting partonic collectivity.

To investigate the underlying physics of the particle–antiparticle $v_2$ splitting, Fig.~\ref{Fig:fig4} shows $v_2$ as a function of $(m_T - m_0)$ for $p$, $\bar{p}$, ${\Lambda}$, $\bar{\Lambda}$, ${\Xi}^{-}$, and $\bar{\Xi}^{+}$ in 10–40\% central Au+Au collisions at $\sqrt{s_{NN}} =$ 11.5 and 9.2 GeV. Panels (a) and (b) show the results at 11.5 and 9.2 GeV, respectively. Particles and antiparticles are represented by filled and open markers, respectively. A clear particle–antiparticle splitting is observed over a broad $(m_T - m_0)$ range.
Remarkably, $p$, $\Lambda$, and $\Xi$ baryons exhibit similar $v_2$ values, and their antiparticles likewise show comparable $v_2$ among themselves, despite the markedly different hadronic cross sections across these species.
Because $\Xi$ experiences weaker hadronic rescattering than $\Lambda$~\cite{Gal:2016boi}, this similarity indicates that hadronic interactions are unlikely to dominate the observed splitting. While hadronic rescattering is known to play an important role in late-stage evolution, the present results suggest that its contribution to the observed particle-antiparticle $v_2$ splitting is subdominant. This comparison indicates that the $v_2$ difference is primarily driven by baryon-related effects, such as baryon transport and possible mean-field potentials~\cite{Steinheimer:2012bn,Xu:2012gf,Xu:2013sta,Liu:2019ags,Dunlop:2011cf}, established at an early stage of the system evolution.

{\bf Summary.} We report high-precision measurements of elliptic flow ($v_2$) for multi-strange hadrons ($\phi$, $\Xi$, and $\Omega$) and light-flavor hadrons ($\pi^{\pm}, K^{\pm}$, $K_{S}^{0}$, $p$, and $\Lambda$) in 10--40\% central Au+Au collisions at $\sqrt{s_{NN}} = 7.7$--19.6 GeV, based on the second phase of the STAR Beam Energy Scan program. 
The $v_2(p_T)$ exhibits mass ordering at low $p_T$ and baryon-meson splitting at intermediate $p_T$, including for $\phi$, $\Xi$, and $\Omega$, demonstrating that collective flow is predominantly established at the partonic stage. 
NCQ scaling is observed to hold for both particles and antiparticles down to $\sqrt{s_{NN}} = 7.7$ GeV, indicating that partonic collectivity persists at these energies. 
Among baryons, $p$, $\Lambda$, and $\Xi$ show similar $v_2$ values, and their antiparticles likewise exhibit comparable $v_2$ among themselves, despite large differences in their hadronic cross sections, providing new constraints beyond BES-I on the origin of the particle-antiparticle $v_2$ splitting and suggesting that the observed differences arise primarily from early-stage baryon-related dynamics, such as baryon transport and mean-field effects, rather than from late-stage hadronic rescattering. 
These results highlight multi-strange hadrons as precision probes of partonic collectivity and support its persistence to $\sqrt{s_{NN}} = 7.7$ GeV, placing new constraints on the onset of quark-gluon plasma-like behavior in the QCD phase diagram.

{\bf Acknowledgments:}
We thank the RHIC Operations Group and RCF at BNL, the NERSC Center at LBNL, and the Open Science Grid consortium for providing resources and support.  This work was supported in part by the Office of Nuclear Physics within the U.S. DOE Office of Science, the U.S. National Science Foundation, the Ministry of Science and Technology of China and the Chinese Ministry of Education, National Natural Science Foundation of China, Chinese Academy of Science, the Higher Education Sprout Project by Ministry of Education at NCKU, the National Research Foundation of Korea, Czech Science Foundation and Ministry of Education, Youth and Sports of the Czech Republic, Hungarian National Research, Development and Innovation Office, New National Excellency Programme of the Hungarian Ministry of Human Capacities, Department of Atomic Energy and Department of Science and Technology of the Government of India, the National Science Centre of Poland, the Ministry of Science, Education and Sports of the Republic of Croatia, German Bundesministerium f\"ur Bildung, Wissenschaft, Forschung and Technologie (BMBF), Helmholtz Association, Ministry of Education, Culture, Sports, Science, and Technology (MEXT) and Japan Society for the Promotion of Science (JSPS).

\bibliography{example} 

@article{Smith:1997xu,
    author = "Smith, Wade and Haglin, Kevin L.",
    title = "{Collision broadening of the phi meson in baryon rich hadronic matter}",
    eprint = "nucl-th/9710026",
    archivePrefix = "arXiv",
    doi = "10.1103/PhysRevC.57.1449",
    journal = "Phys. Rev. C",
    volume = "57",
    pages = "1449--1453",
    year = "1998"
}

@article{Chung:1997mp,
    author = "Chung, W. S. and Li, Guo-Qiang and Ko, C. M.",
    title = "{Phi meson production in heavy ion collisions at SIS energies}",
    eprint = "nucl-th/9704002",
    archivePrefix = "arXiv",
    doi = "10.1016/S0375-9474(97)00198-X",
    journal = "Nucl. Phys. A",
    volume = "625",
    pages = "347--371",
    year = "1997"
}

@article{Ferrari:1959,
  author         = "Ferrari, F. and Fonda, L.",
  title          = "{On the Hyperon-Nucleon Scattering and Reaction Cross Sections}",
  journal        = "Phys. Rev.",
  volume         = "114",
  pages          = "874",
  year           = "1959",
  doi            = "10.1103/PhysRev.114.874"
}

@article{Gal:2016boi,
    author = "Gal, A. and Hungerford, E. V. and Millener, D. J.",
    title = "{Strangeness in nuclear physics}",
    eprint = "1605.00557",
    archivePrefix = "arXiv",
    primaryClass = "nucl-th",
    doi = "10.1103/RevModPhys.88.035004",
    journal = "Rev. Mod. Phys.",
    volume = "88",
    number = "3",
    pages = "035004",
    year = "2016"
}

@article{Crede:2024hur,
    author = "Crede, Volker and Yelton, John",
    title = "{70 years of hyperon spectroscopy: a review of strange $\Xi$, $\Omega$ baryons, and the spectrum of charmed and bottom baryons}",
    eprint = "2502.08815",
    archivePrefix = "arXiv",
    primaryClass = "hep-ex",
    doi = "10.1088/1361-6633/ad7610",
    journal = "Rept. Prog. Phys.",
    volume = "87",
    number = "10",
    pages = "106301",
    year = "2024"
}

@article{Wang:2023vqi,
    author = "Wang, K. and Zhou, J. and Wang, X. and Li, X. and Hu, D. and Sun, Y.",
    title = "{Batch testing and noise rate optimization of MRPC3b for CBM-TOF/STAR-eTOF}",
    doi = "10.1016/j.nima.2023.168778",
    journal = "Nucl. Instrum. Meth. A",
    volume = "1057",
    pages = "168778",
    year = "2023",
    eprint       = {2308.16556},
    archivePrefix= {arXiv},
    primaryClass = {physics.ins-det}
}

@article{Yang:2017llt,
    title = {The STAR beam energy scan phase II physics and upgrades},
    journal = {Nuclear Physics A},
    volume = {967},
    pages = {800-803},
    year = {2017},
    issn = {0375-9474},
    doi = {https://doi.org/10.1016/j.nuclphysa.2017.05.042},
    url = {https://www.sciencedirect.com/science/article/pii/S0375947417301598},
    author = {Chi Yang},
	collaboration = {STAR Collaboration},
    eprint       = {1810.04767},
    archivePrefix= {arXiv},
    primaryClass = {nucl-ex}
}

@article{Adams:2019fpo,
	eprint       = {1912.05243},
    archivePrefix= {arXiv},
	author = {Adams, Joseph and others},
	doi = {10.1016/j.nima.2020.163970},
	journal = {Nucl. Instrum. Meth. A},
	pages = {163970},
	primaryclass = {physics.ins-det},
	title = "{The STAR Event Plane Detector}",
	volume = {968},
	year = {2020}}

@article{Miller:2007ri,
    author = "Miller, Michael L. and Reygers, Klaus and Sanders, Stephen J. and Steinberg, Peter",
    title = "{Glauber modeling in high energy nuclear collisions}",
    eprint = "nucl-ex/0701025",
    archivePrefix = "arXiv",
    doi = "10.1146/annurev.nucl.57.090506.123020",
    journal = "Ann. Rev. Nucl. Part. Sci.",
    volume = "57",
    pages = "205--243",
    year = "2007"
}

@article{Kisel:2018nvd,
	author = {Kisel, Ivan},
	collaboration = {CBM Collaboration},
	doi = {10.1088/1742-6596/1070/1/012015},
	editor = {Aichelin, Joerg and Bellwied, Rene and Bratkovskaya, Elena and Timmins, Anthony},
	journal = {J. Phys. Conf. Ser.},
	number = {1},
	pages = {012015},
	title = {{Event Topology Reconstruction in the CBM Experiment}},
	volume = {1070},
	year = {2018}}

@article{Banerjee:2020iab,
	author = {Banerjee, Arundhati and Kisel, Ivan and Zyzak, Maksym},
	doi = {10.1142/S0217751X20430034},
	journal = {Int. J. Mod. Phys. A},
	number = {33},
	pages = {2043003},
	title = {{Artificial neural network for identification of short-lived particles in the CBM experiment}},
	volume = {35},
	year = {2020}}

@article{Voloshin:2008dg,
    author = "Voloshin, Sergei A. and Poskanzer, Arthur M. and Snellings, Raimond",
    editor = "Stock, R.",
    title = "{Collective phenomena in non-central nuclear collisions}",
    eprint = "0809.2949",
    archivePrefix = "arXiv",
    primaryClass = "nucl-ex",
    doi = "10.1007/978-3-642-01539-7_10",
    journal = "Landolt-Bornstein",
    volume = "23",
    pages = "293--333",
    year = "2010"
}

@article{Dong_2004,
   title={Resonance decay effects on anisotropy parameters},
   volume={597},
   ISSN={0370-2693},
   url={http://dx.doi.org/10.1016/j.physletb.2004.06.110},
   DOI={10.1016/j.physletb.2004.06.110},
   number={3–4},
   journal={Physics Letters B},
   publisher={Elsevier BV},
   author={Dong, X. and Esumi, S. and Sorensen, P. and Xu, N. and Xu, Z.},
   year={2004},
   month=sep, pages={328–332},
   eprint       = {nucl-th/0403030},
  archivePrefix= {arXiv},
   }

@article{STAR:2013cow,
    author = "Adamczyk, L. and others",
    collaboration = "STAR Collaboration",
    title = "{Observation of an Energy-Dependent Difference in Elliptic Flow between Particles and Antiparticles in Relativistic Heavy Ion Collisions}",
    eprint = "1301.2347",
    archivePrefix = "arXiv",
    primaryClass = "nucl-ex",
    doi = "10.1103/PhysRevLett.110.142301",
    journal = "Phys. Rev. Lett.",
    volume = "110",
    number = "14",
    pages = "142301",
    year = "2013"
}

@article{YONG2021136521,
title = {Double strangeness Ξ− production as a probe of nuclear equation of state at high densities},
journal = {Physics Letters B},
volume = {820},
pages = {136521},
year = {2021},
issn = {0370-2693},
doi = {https://doi.org/10.1016/j.physletb.2021.136521},
url = {https://www.sciencedirect.com/science/article/pii/S0370269321004615},
author = {Gao-Chan Yong and Zhi-Gang Xiao and Yuan Gao and Zi-Wei Lin},
eprint = "2105.10284",
archivePrefix = "arXiv",
primaryClass = "nucl-th"
}

@article{Chen:2024BESreview,
  author = "Chen, Jin-Hui and Dong, Xin and He, Xiong-Hong and Huang, Huan-Zhong and Liu, Feng and Luo, Xiao-Feng and Ma, Yu-Gang and et al.",
  title = "{Properties of the QCD matter: review of selected results from the relativistic heavy ion collider beam energy scan (RHIC BES) program}",
  journal = "Nucl. Sci. Tech.",
  volume = "35",
  pages = "214",
  year = "2024",
  doi = "10.1007/s41365-024-01591-2",
  eprint = "2407.02935",
  archivePrefix = "arXiv",
  primaryClass = "nucl-ex"
}

@article{STAR:2013ayu,
    author = "Adamczyk, L. and others",
    collaboration = "STAR Collaboration",
    title = "{Elliptic flow of identified hadrons in Au+Au collisions at $\sqrt{s_{\mathrm{NN}}}=$ 7.7-62.4 GeV}",
    eprint = "1301.2348",
    archivePrefix = "arXiv",
    primaryClass = "nucl-ex",
    doi = "10.1103/PhysRevC.88.014902",
    journal = "Phys. Rev. C",
    volume = "88",
    pages = "014902",
    year = "2013"
}

@article{STAR:2015rxv,
    author = "Adamczyk, L. and others",
    collaboration = "STAR Collaboration",
    title = "{Centrality dependence of identified particle elliptic flow in relativistic heavy ion collisions at $\sqrt{s_{\mathrm{NN}}}=$ 7.7-62.4 GeV}",
    eprint = "1509.08397",
    archivePrefix = "arXiv",
    primaryClass = "nucl-ex",
    doi = "10.1103/PhysRevC.93.014907",
    journal = "Phys. Rev. C",
    volume = "93",
    number = "1",
    pages = "014907",
    year = "2016"
}

@article{STAR:2012och,
    author = "Adamczyk, L. and others",
    collaboration = "STAR",
    title = "{Inclusive charged hadron elliptic flow in Au + Au collisions at $\sqrt{s_{\mathrm{NN}}}=$ 7.7-39 GeV}",
    eprint = "1206.5528",
    archivePrefix = "arXiv",
    primaryClass = "nucl-ex",
    doi = "10.1103/PhysRevC.86.054908",
    journal = "Phys. Rev. C",
    volume = "86",
    pages = "054908",
    year = "2012"
}

@article{STAR:2025owm,
    author = "Aboona, B. E. and others",
    collaboration = "STAR Collaboration",
    title = "{Onset of Constituent Quark Number Scaling in Heavy-Ion Collisions at RHIC}",
    eprint = "2504.02531",
    archivePrefix = "arXiv",
    primaryClass = "nucl-ex",
    doi = "10.1103/2qhx-cp79",
    journal = "Phys. Rev. Lett.",
    volume = "135",
    number = "7",
    pages = "072301",
    year = "2025"
}

@article{STAR:2010vob,
    author = "Aggarwal, M. M. and others",
    collaboration = "STAR Collaboration",
    title = "{An Experimental Exploration of the QCD Phase Diagram: The Search for the Critical Point and the Onset of De-confinement}",
    eprint = "1007.2613",
    archivePrefix = "arXiv",
    primaryClass = "nucl-ex",
    month = "7"
}

@misc{SupplementalMaterial,
  note = {See Supplemental Material at \url{http://......} for details}
}

@article{SHEN201890,
  author = {Fuwang Shen and others},
title = {MWPC prototyping and performance test for the STAR inner TPC upgrade},
journal = {Nucl. Instrum. Meth. A},
volume = {896},
pages = {90-95},
year = {2018},
issn = {0168-9002},
doi = {https://doi.org/10.1016/j.nima.2018.04.019}
}

@article{PHENIX:2006dpn,
    author = "Adare, A. and others",
    collaboration = "PHENIX Collaboration",
    title = "{Scaling properties of azimuthal anisotropy in Au+Au and Cu+Cu collisions at $\sqrt{s_{\mathrm{NN}}}=$ 200 GeV}",
    eprint = "nucl-ex/0608033",
    archivePrefix = "arXiv",
    doi = "10.1103/PhysRevLett.98.162301",
    journal = "Phys. Rev. Lett.",
    volume = "98",
    pages = "162301",
    year = "2007"
}

@article{ALICE:2014wao,
    author = "Abelev, Betty Bezverkhny and others",
    collaboration = "ALICE",
    title = "{Elliptic flow of identified hadrons in Pb-Pb collisions at $ \sqrt{s_{\mathrm{NN}}}=$ 2.76 TeV}",
    eprint = "1405.4632",
    archivePrefix = "arXiv",
    primaryClass = "nucl-ex",
    reportNumber = "CERN-PH-EP-2014-104",
    doi = "10.1007/JHEP06(2015)190",
    journal = "JHEP",
    volume = "06",
    pages = "190",
    year = "2015"
}

@article{Poskanzer:1998yz,
  author = {Poskanzer, A. M. and Voloshin, S. A.},
  title = {Methods for analyzing anisotropic flow in relativistic nuclear collisions},
  journal = {Phys. Rev. C},
  volume = {58},
  pages = {1671--1678},
  year = {1998},
  eprint = "nucl-ex/9805001",
  archivePrefix = "arXiv",
  doi = {10.1103/PhysRevC.58.1671}
  
}

@article{STAR:2003wqp,
    author = "Adams, John and others",
    collaboration = "STAR Collaboration",
    title = "{Particle type dependence of azimuthal anisotropy and nuclear modification of particle production in Au + Au collisions at $\sqrt{s_{\mathrm{NN}}}=$ 200 GeV}",
    eprint = "nucl-ex/0306007",
    archivePrefix = "arXiv",
    doi = "10.1103/PhysRevLett.92.052302",
    journal = "Phys. Rev. Lett.",
    volume = "92",
    pages = "052302",
    year = "2004"
}

@article{Adams:2005dq,
  author       = {J. Adams and others},
  collaboration = {STAR Collaboration},
  title        = "{Multistrange Baryon Elliptic Flow in Au+Au Collisions at $\sqrt{s_{\mathrm{NN}}}=$ 200 GeV}",
  journal      = {Phys. Rev. Lett.},
  volume       = {95},
  pages        = {122301},
  year         = {2005},
  doi          = {10.1103/PhysRevLett.95.122301},
  eprint       = {nucl-ex/0504022},
  archivePrefix= {arXiv}
}

@article{Abelev:2007rw,
  author        = {B. I. Abelev and others},
  collaboration = {STAR Collaboration},
  title         = {Mass, quark-number, and $\sqrt{s_{\mathrm{NN}}}$ dependence of the second and fourth flow harmonics in ultra-relativistic nucleus-nucleus collisions},
  journal       = {Phys. Rev. C},
  volume        = {75},
  pages         = {054906},
  year          = {2007},
  doi           = {10.1103/PhysRevC.75.054906},
  eprint        = {nucl-ex/0701010},
  archivePrefix = {arXiv}
}

@article{STAR:2015gge,
    author = "Adamczyk, L. and others",
    collaboration = "STAR Collaboration",
    title = "{Centrality and transverse momentum dependence of elliptic flow of multistrange hadrons and $\phi$ meson in Au+Au collisions at $\sqrt{s_{\mathrm{NN}}}=$ 200 GeV}",
    eprint = "1507.05247",
    archivePrefix = "arXiv",
    primaryClass = "nucl-ex",
    doi = "10.1103/PhysRevLett.116.062301",
    journal = "Phys. Rev. Lett.",
    volume = "116",
    number = "6",
    pages = "062301",
    year = "2016"
}

@article{Adamczyk:2017xur,
  author        = {L. Adamczyk and others},
  collaboration = {STAR Collaboration},
  title         = "{Measurement of $D^0$ Azimuthal Anisotropy at Midrapidity in Au+Au Collisions at $\sqrt{s_{\mathrm{NN}}}=$ 200 GeV}",
  journal       = {Phys. Rev. Lett.},
  volume        = {118},
  pages         = {212301},
  year          = {2017},
  doi           = {10.1103/PhysRevLett.118.212301},
  eprint        = {1701.06060},
  archivePrefix = {arXiv},
  primaryClass  = {nucl-ex}
}

@article{Abelev:2007qg,
  author       = {B. I. Abelev and others},
  collaboration = {STAR Collaboration},
  title        = "{Centrality dependence of charged hadron and strange hadron elliptic flow from $\sqrt{s_{\mathrm{NN}}}=$ 200 GeV Au+Au collisions}",
  journal      = {Phys. Rev. C},
  volume       = {77},
  pages        = {054901},
  year         = {2008},
  doi          = {10.1103/PhysRevC.77.054901},
  eprint       = {0801.3466},
  archivePrefix= {arXiv},
  primaryClass = {nucl-ex}
}

@article{STAR:2008bgi,
    author = "Abelev, B. I. and others",
    collaboration = "STAR Collaboration",
    title = "{Measurements of phi meson production in relativistic heavy-ion collisions at RHIC}",
    eprint = "0809.4737",
    archivePrefix = "arXiv",
    primaryClass = "nucl-ex",
    doi = "10.1103/PhysRevC.79.064903",
    journal = "Phys. Rev. C",
    volume = "79",
    pages = "064903",
    year = "2009"
}

@article{Masui:2012zh,
    author = "Masui, Hiroshi and Schmah, Alexander and Poskanzer, A. M.",
    title = "{Event plane resolution correction for azimuthal anisotropy in wide centrality bins}",
    eprint = "1212.3650",
    archivePrefix = "arXiv",
    primaryClass = "physics.data-an",
    doi = "10.1016/j.nima.2016.07.037",
    journal = "Nucl. Instrum. Meth. A",
    volume = "833",
    pages = "181--185",
    year = "2016"
}

@article{Barlow:2002yb,
    title={Systematic Errors: facts and fictions}, 
      author={Roger Barlow},
      eprint={hep-ex/0207026},
      archivePrefix={arXiv},
      url={https://arxiv.org/abs/hep-ex/0207026}
}

@article{Steinheimer:2012bn,
    author = "Steinheimer, J. and Koch, V. and Bleicher, M.",
    title = "{Hydrodynamics at large baryon densities: Understanding proton vs. anti-proton $v_2$ and other puzzles}",
    eprint = "1207.2791",
    archivePrefix = "arXiv",
    primaryClass = "nucl-th",
    doi = "10.1103/PhysRevC.86.044903",
    journal = "Phys. Rev. C",
    volume = "86",
    pages = "044903",
    year = "2012"
}

@article{Xu:2013sta,
    author = "Xu, Jun and Song, Taesoo and Ko, Che Ming and Li, Feng",
    title = "{Elliptic flow splitting as a probe of the QCD phase structure at finite baryon chemical potential}",
    eprint = "1308.1753",
    archivePrefix = "arXiv",
    primaryClass = "nucl-th",
    doi = "10.1103/PhysRevLett.112.012301",
    journal = "Phys. Rev. Lett.",
    volume = "112",
    pages = "012301",
    year = "2014"
}

@article{Liu:2019ags,
    author = "Liu, He and Wang, Feng-Tao and Sun, Kai-Jia and Xu, Jun and Ko, Che Ming",
    title = "{Isospin splitting of pion elliptic flow in relativistic heavy-ion collisions}",
    eprint = "1908.01156",
    archivePrefix = "arXiv",
    primaryClass = "nucl-th",
    doi = "10.1016/j.physletb.2019.135002",
    journal = "Phys. Lett. B",
    volume = "798",
    pages = "135002",
    year = "2019"
}

@article{Dunlop:2011cf,
    author = "Dunlop, J. C. and Lisa, M. A. and Sorensen, P.",
    title = "{Constituent quark scaling violation due to baryon number transport}",
    eprint = "1107.3078",
    archivePrefix = "arXiv",
    primaryClass = "hep-ph",
    doi = "10.1103/PhysRevC.84.044914",
    journal = "Phys. Rev. C",
    volume = "84",
    pages = "044914",
    year = "2011"
}

@article{Voloshin:2005up,
	author = {Voloshin, S. A.},
	doi = {10.1088/1742-6596/9/1/052},
	editor = {Barnes, Ted and Godfrey, Steve and Petrov, Alexey A. and Swanson, Eric},
	journal = {J. Phys. Conf. Ser.},
	pages = {276--279},
	title = {{Anisotropic flow at RHIC: Constituent quark scaling}},
	volume = {9},
	year = {2005}}

@article{CMS:2018loe,
    author = "Sirunyan, A. M. and others",
    collaboration = "CMS Collaboration",
    title = "{Elliptic flow of charm and strange hadrons in high-multiplicity pPb collisions at $\sqrt{s_{_\mathrm{NN}}} =$ 8.16 TeV}",
    eprint = "1804.09767",
    archivePrefix = "arXiv",
    primaryClass = "hep-ex",
    reportNumber = "CMS-HIN-17-003, CERN-EP-2018-076",
    doi = "10.1103/PhysRevLett.121.082301",
    journal = "Phys. Rev. Lett.",
    volume = "121",
    number = "8",
    pages = "082301",
    year = "2018"
}

@article{Chen:2026gka,
    author = "Chen, Jinhui and others",
    title = "{Selected Highlights from STAR Experiment}",
    eprint = "2601.12977",
    archivePrefix = "arXiv",
    primaryClass = "nucl-ex",
    doi = "10.1088/0256-307X/43/3/030102",
    journal = "Chin. Phys. Lett.",
    volume = "43",
    number = "3",
    pages = "030102",
    year = "2026"
}

@article{Denicol:2018wdp,
    author = {Denicol, Gabriel S. and Gale, Charles and Jeon, Sangyong and Monnai, Akihiko and Schenke, Bj\"orn and Shen, Chun},
    title = "{Net baryon diffusion in fluid dynamic simulations of relativistic heavy-ion collisions}",
    eprint = "1804.10557",
    archivePrefix = "arXiv",
    primaryClass = "nucl-th",
    reportNumber = "KEK-TH-2075",
    doi = "10.1103/PhysRevC.98.034916",
    journal = "Phys. Rev. C",
    volume = "98",
    number = "3",
    pages = "034916",
    year = "2018"
}

@article{Bilandzic:2010jr,
    author = "Bilandzic, Ante and Snellings, Raimond and Voloshin, Sergei",
    title = "{Flow analysis with cumulants: Direct calculations}",
    eprint = "1010.0233",
    archivePrefix = "arXiv",
    primaryClass = "nucl-ex",
    doi = "10.1103/PhysRevC.83.044913",
    journal = "Phys. Rev. C",
    volume = "83",
    pages = "044913",
    year = "2011"
}

@article{He:2017tla,
    author = "He, Yuncun and Lin, Zi-Wei",
    title = "{Improved Quark Coalescence for a Multi-Phase Transport Model}",
    eprint = "1703.02673",
    archivePrefix = "arXiv",
    primaryClass = "nucl-th",
    doi = "10.1103/PhysRevC.96.014910",
    journal = "Phys. Rev. C",
    volume = "96",
    number = "1",
    pages = "014910",
    year = "2017"
}

@article{Huovinen:2001cy,
    author = "Huovinen, P. and Kolb, P. F. and Heinz, Ulrich W. and Ruuskanen, P. V. and Voloshin, S. A.",
    title = "{Radial and elliptic flow at RHIC: Further predictions}",
    eprint = "hep-ph/0101136",
    archivePrefix = "arXiv",
    doi = "10.1016/S0370-2693(01)00219-2",
    journal = "Phys. Lett. B",
    volume = "503",
    pages = "58--64",
    year = "2001"
}

@article{Moln_r_2003,
   title={Elliptic Flow at Large Transverse Momenta from Quark Coalescence},
   volume={91},
   ISSN={1079-7114},
   number={9},
   journal={Phys. Rev. Lett.},
   publisher={American Physical Society (APS)},
   author={Molnár, Dénes and Voloshin, Sergei A.},
   year={2003},
   eprint       = {nucl-th/0302014},
   archivePrefix= {arXiv},
   url={http://dx.doi.org/10.1103/PhysRevLett.91.092301},
   month="aug"
   }

@article{STAR:2007mum,
    author = "Abelev, B. I. and others",
    collaboration = "STAR Collaboration",
    title = "{Partonic flow and phi-meson production in Au + Au collisions at $\sqrt{s_{\mathrm{NN}}}=$ 200 GeV}",
    eprint = "nucl-ex/0703033",
    archivePrefix = "arXiv",
    doi = "10.1103/PhysRevLett.99.112301",
    journal = "Phys. Rev. Lett.",
    volume = "99",
    pages = "112301",
    year = "2007"
}

@article{Xu:2012gf,
    author = "Xu, Jun and Chen, Lie-Wen and Ko, Che Ming and Lin, Zi-Wei",
    title = "{Effects of hadronic potentials on elliptic flows in relativistic heavy ion collisions}",
    eprint = "1201.3391",
    archivePrefix = "arXiv",
    primaryClass = "nucl-th",
    doi = "10.1103/PhysRevC.85.041901",
    journal = "Phys. Rev. C",
    volume = "85",
    pages = "041901",
    year = "2012"
}

@article{ParticleDataGroup:2024cfk,
    author = "Navas, S. and others",
    collaboration = "Particle Data Group",
    title = "{Review of particle physics}",
    doi = "10.1103/PhysRevD.110.030001",
    journal = "Phys. Rev. D",
    volume = "110",
    number = "3",
    pages = "030001",
    year = "2024"
}

\end{document}



\title{Supplemental Material: Elliptic Flow of Multi-Strange Hadrons in Au+Au Collisions at $\sqrt{s_{NN}}$ = 7.7–19.6 GeV} 

\bigskip

\author{The STAR Collaboration}
\date{\today}
\begin{abstract}
\end{abstract}

\maketitle
\onecolumngrid

\section{$\phi$-meson and strange hadrons Invariant mass distributions}

\begin{figure*}[htbp]
  \includegraphics[width=0.96\columnwidth]{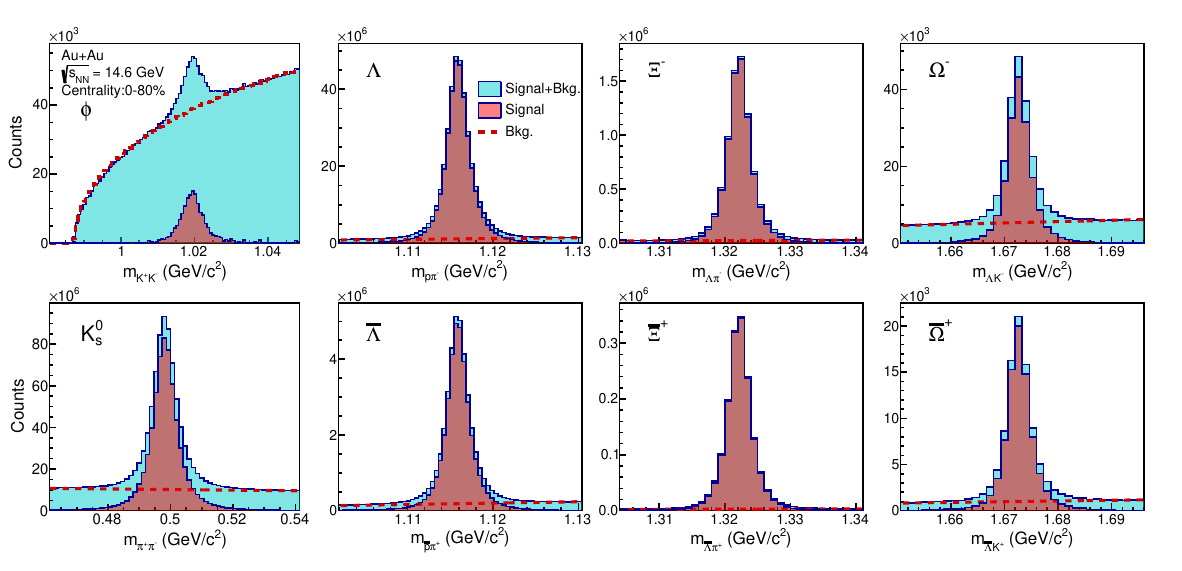}
  \caption{Invariant mass distributions of $\phi$, $K^{0}_{S}$, $\Lambda$, $\bar{\Lambda}$, $\Xi^{\pm}$, and $\Omega^{\pm}$ in 0–80\% centrality Au+Au collisions at $\rootsNN$ = 14.6 GeV.}     
  \label{Fig:fig_particles_inv}
\end{figure*}

Figure~\ref{Fig:fig_particles_inv} shows the invariant mass distributions of $\phi$, $K^{0}_{S}$, $\Lambda$, $\bar{\Lambda}$, $\Xi^{\pm}$, and $\Omega^{\pm}$ in 0--80\% Au+Au collisions at $\rootsNN = 14.6$ GeV. 
The blue shaded region represents the invariant mass distribution reconstructed from the daughter track pairs. 
The combinatorial background, estimated using the mixed-event method for $\phi$ mesons and linear fits to accommodate residual background for the other strange particles, is shown by the red dashed lines. 
The signal after background subtraction is displayed in the red shaded area.

\section{Transverse momentum dependence of elliptic flow}

\begin{figure*}[!htb]
    \centering
    \includegraphics[width=0.7\textwidth]{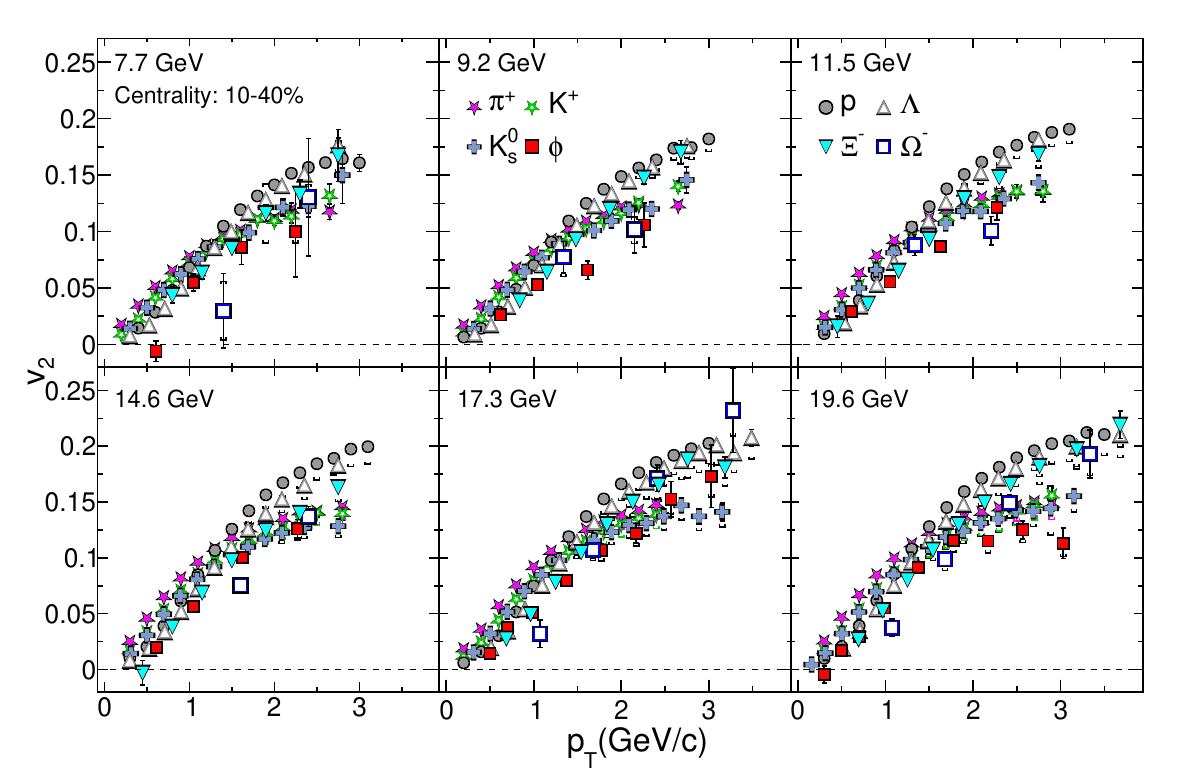}
    \caption{The elliptic flow as a function of transverse momentum, $v_2$($p_T$), in 10--40\% central Au+Au collisions for selected particles at {\rootsNN} = 7.7 -- 19.6 GeV. The error bars and brackets represent statistical and systematic uncertainties, respectively.}
    \label{Fig:fig_v2pt1}
\end{figure*}

\begin{figure*}[!htb]
    \centering
    \includegraphics[width=0.7\textwidth]{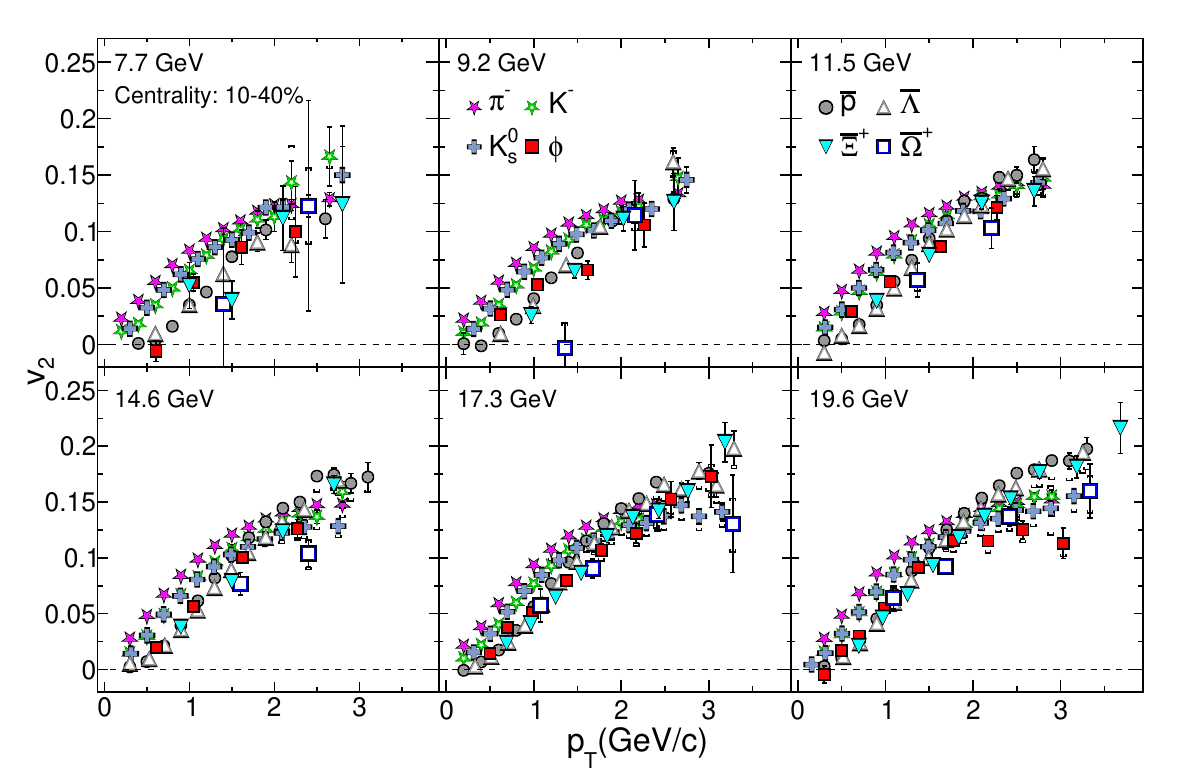}
    \caption{The elliptic flow as a function of transverse momentum, $v_2$($p_T$), in 10--40\% central Au+Au collisions for selected antiparticles at {\rootsNN} = 7.7 -- 19.6 GeV. The error bars and brackets represent statistical and systematic uncertainties, respectively.}
    \label{Fig:fig_v2pt2}
\end{figure*}

Figures ~\ref{Fig:fig_v2pt1} and ~\ref{Fig:fig_v2pt2} present the transverse momentum ($p_T$) dependence of the elliptic flow ($v_2$) for  ${\pi}^{\pm}$, ${K}^{\pm}$, ${K}^{0}_{S}$, $p$, $\bar{p}$, ${\phi}$, ${\Lambda}$, $\bar{\Lambda}$, ${\Xi}^{\pm}$, and ${\Omega}^{\pm}$ in 10--40\% central Au+Au collisions at {\rootsNN} = 7.7 - 19.6 GeV. For completeness, the figures present the results for all measured particle species, including the pion and kaon results that are not shown in the main text.
The former presents the results for particles, while the latter shows those for antiparticles. 
For all beam energies and all particle species, a clear mass ordering is observed at low $p_T$: lighter particles show larger $v_2$ values than heavier ones, which is consistent with the expectation from hydrodynamic collective expansion~\cite{Huovinen:2001cy}. 

At intermediate $p_T$, a clear baryon-meson separation is observed, where baryons exhibit systematically larger $v_2$ values than mesons. This feature is naturally explained by the quark coalescence mechanism, in which hadrons inherit the collective flow of their constituent quarks. The observation provides strong evidence that partonic collectivity has been developed and that partonic degrees of freedom dominate the medium of Au + Au collisions at $\rootsNN \ge 7.7$ GeV~\cite{Moln_r_2003,STAR:2003wqp,STAR:2007mum}.

\section{Number of Constituent Quarks (NCQ) scaling: $v_{2}/n_{q}$ vs. $p_{T}/n_{q}$}

\begin{figure*}[!htb]
    \centering
    \includegraphics[width=0.8\textwidth]{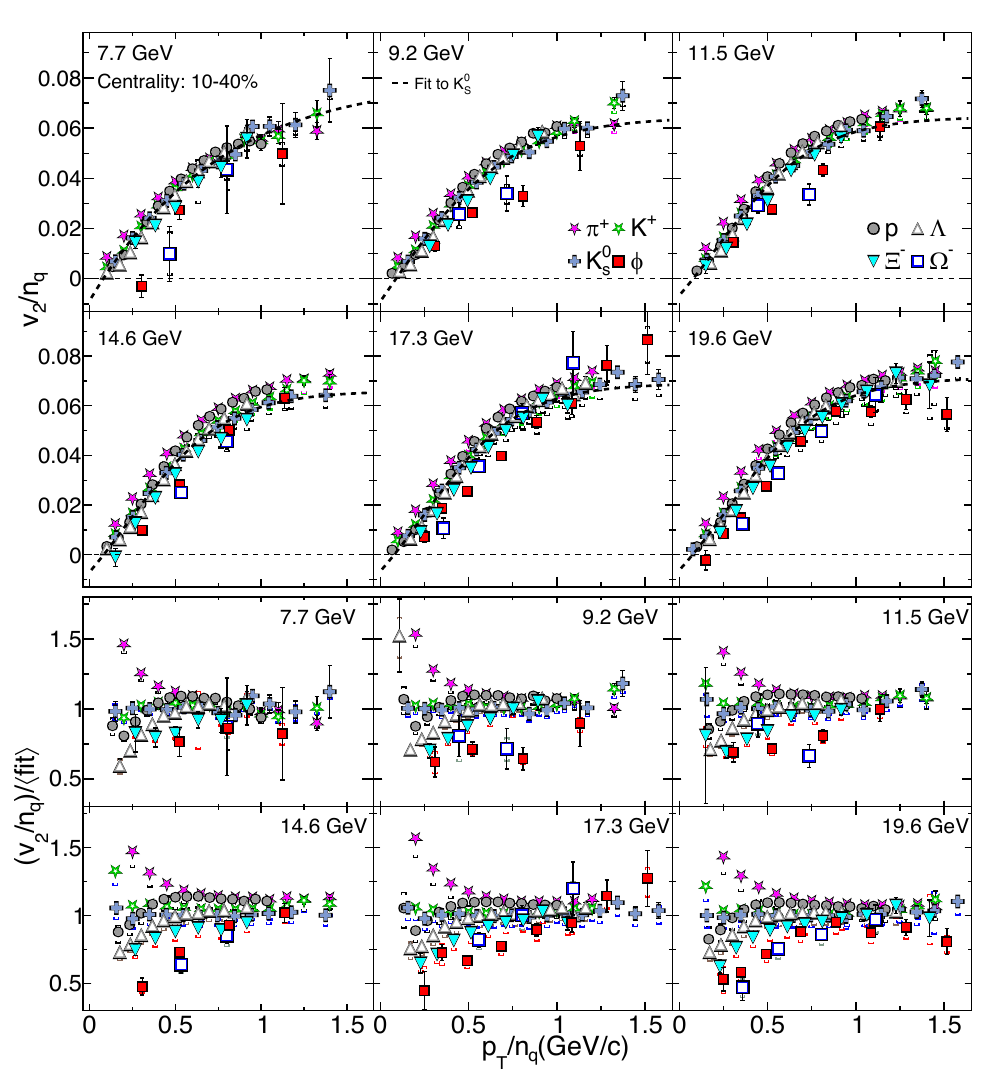}
    \caption{The NCQ-scaled elliptic flow, $v_2/n_q$ versus $p_T/n_q$, is shown for particles in 10--40\% centrality Au+Au collisions, along with the NCQ scaling ratios relative to the $K_S^0$ fit function~\cite{Dong_2004}. The error bars and brackets represent statistical and systematic uncertainties, respectively.}
    \label{Fig:fig_NCQ_pT_nq1}
\end{figure*}

\begin{figure*}[!htb]
    \centering
    \includegraphics[width=0.8\textwidth]{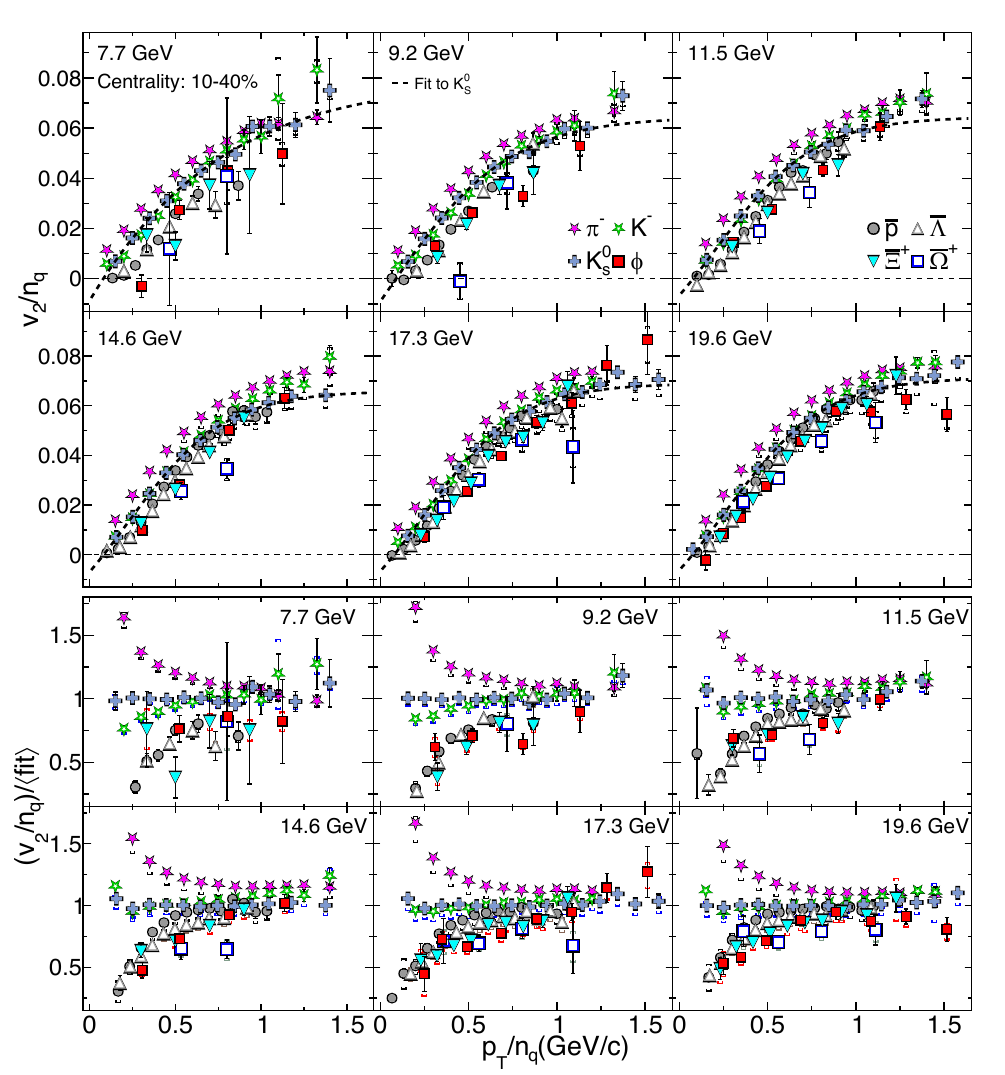}
    \caption{The NCQ-scaled elliptic flow, $v_2/n_q$ versus $p_T/n_q$, is shown for antiparticles in 10--40\% centrality Au+Au collisions, along with the NCQ scaling ratios relative to the $K_S^0$ fit function~\cite{Dong_2004}. The error bars and brackets represent statistical and systematic uncertainties, respectively.}
    \label{Fig:fig_NCQ_pT_nq2}
\end{figure*}

Figures~\ref{Fig:fig_NCQ_pT_nq1} and ~\ref{Fig:fig_NCQ_pT_nq2} represent the number of quarks ($n_q$) scaled \(v_2\) as a function of \(p_T/n_q\) for particles and antiparticles in 10-40\% central Au+Au collisions, respectively. 
At low \(p_T/n_q\), the scaled $v_2$ values of different hadrons are separated from each other because of the hydrodynamic mass ordering effect driven by radial flow~\cite{Huovinen:2001cy}. 
In comparison, when the same data are plotted as $v_2/n_q$ versus the scaled transverse kinetic energy ($(m_T - m_0)/n_q$), where $m_T = \sqrt{p^{2}_{T}+m^{2}_0}$, $m_0$ is the rest mass of the particle, the separation becomes smaller, and the scaling among different hadrons appears improved. This improvement arises because the use of $(m_T - m_0)$ partially accounts for the particle mass dependence of the transverse motion, which reduces the influence of radial flow on the scaling behavior~\cite{PHENIX:2006dpn}.

As \(p_T/n_q \gtrsim 1\)~GeV/\(c\), the approximate NCQ scaling holds, and baryons and mesons follow a common trend, indicating quark-level collectivity. 
Notably, the multi-strange hadrons ($\Xi^{\pm}$, $\Omega^{\pm}$) and the ${\phi}$ meson also follow this common trend, despite their relatively small hadronic cross sections, further supporting the picture that the observed collectivity originates at the partonic level when $\rootsNN \ge 7.7$ GeV.

\bibliography{example} 